%% file: letter.tex
\documentclass[aps, prl, reprint, longbibliography, floatfix, nobibnotes, superscriptaddress]{revtex4-2}

\usepackage{aas_macros}
\usepackage{epsfig}
\usepackage{url}
\usepackage{hyperref}

\usepackage[normalem]{ulem} 

\usepackage{latexsym}
\usepackage{epsfig}
\usepackage{amsmath}
\usepackage{amssymb}
\usepackage{wasysym}
\usepackage{graphicx}
\usepackage{dcolumn}
\usepackage{verbatim}
\usepackage{enumerate,mdwlist}
\usepackage[titletoc]{appendix}
\usepackage{amsfonts}
\usepackage{fancyvrb} 
\usepackage{figchild}
\usepackage{tikz} 
\usetikzlibrary{calc}
\usepackage[export]{adjustbox}

\usepackage{float}
\usepackage{booktabs}
\usepackage{siunitx}   
\usepackage{galdrastafir}

\usepackage[normalem]{ulem}

\newcommand{\glow}{\texttt{GLoW}}
\newcommand{\lettersection}[1]{\textbf{\emph{#1}}.---}
\newcommand{\this}{\emph{Letter}}

\usepackage{xspace}

\newcommand{\go}{\gls{GO}\xspace}
\newcommand{\wo}{\gls{WO}\xspace}
\newcommand{\gws}{\glspl{GW}\xspace}
\newcommand{\cdm}{\gls{CDM}\xspace}
\newcommand{\wdm}{\gls{WDM}\xspace}
\newcommand{\sidm}{\gls{SIDM}\xspace}
\newcommand{\dm}{\gls{DM}\xspace}

\newcommand{\tnfw}{\gls{TNFW}\xspace}
\newcommand{\tnfwc}{\gls{TNFWC}\xspace}

\definecolor{grey}{rgb}{0.4,0.4,0.4}
\definecolor{dullmagenta}{rgb}{0.4,0,0.4}
\definecolor{darkblue}{rgb}{0,0,0.4}
\definecolor{midblue}{rgb}{0,0,0.5}
\definecolor{midred}{rgb}{0.5,0,0}
\definecolor{orange}{rgb}{1,0.5,0}
\definecolor{lightbrown}{rgb}{0.75,0.5,0.25}
\definecolor{tan}{cmyk}{0.14,0.42,0.56,0}
\definecolor{djunglegreen}{cmyk}{0.99,0,0.52,0}
\definecolor{lightgreen}{rgb}{0,1,0}
\definecolor{olivegreen}{cmyk}{0.64,0,0.95,0.40}
\definecolor{midgreen}{rgb}{0.0,0.675,0.0}
\definecolor{darkgreen}{rgb}{0,0.5,0}
\definecolor{ceruleanblue}{rgb}{0.0, 0.2, 0.7}
\definecolor{burgundy}{rgb}{0.5, 0.0, 0.13}
\definecolor{oxblood}{rgb}{0.5333, 0.0314, 0.0314}
\definecolor{steelblue}{rgb}{0.70,0.130,0.180}
\definecolor{maddiblue}{HTML}{4682B4}

\definecolor{hvred}{RGB}{186,12,47}

\hypersetup{
    colorlinks=true,
    linkcolor=oxblood,
    filecolor=oxblood,      
    urlcolor=oxblood,
    citecolor=oxblood,
}

\input{acronyms}

\usepackage[all]{xy} 
\usepackage{amsfonts}
\providecommand{\selectlanguage}[1]{}

\usepackage{staves}

\newcommand{\marksize}{\fontsize{3.4}{4.4}\selectfont}
\makeatletter
\def\@fnsymbol#1{%
  \ifcase#1\or
    \hulinhjalmur[size=1.em]\or
    \marksize\staveXXXV\or 
    \marksize\staveI\or
    \staveII\or
    \staveIII\or
    \staveIV\or
    \staveV\or
    \staveVI\or
    \staveVII\or
    \staveVIII\or
    \staveIX\else
    \@ctrerr
  \fi
}
\makeatother

\makeatletter
\newcommand*{\glsplainhyperlink}[2]{%
  \colorlet{currenttext}{.}
  \colorlet{currentlink}{\@linkcolor}
  \hypersetup{linkcolor=currenttext}
  \hyperlink{#1}{#2}%
  \hypersetup{linkcolor=currentlink}
}
\let\@glslink\glsplainhyperlink
\def\l@subsubsection#1#2{}
\makeatother

\begin{document}

\title{Constraining the Nature of Dark Matter Subhalos with Lensed Gravitational Waves}

\author{Luka Vujeva}
\email{luka.vujeva@nbi.ku.dk}
\affiliation{Center of Gravity, Niels Bohr Institute, Blegdamsvej 17, 2100 Copenhagen, Denmark}
\author{Maddalena Reggio}
\email{maddalena.reggio@nbi.ku.dk}
\affiliation{Center of Gravity, Niels Bohr Institute, Blegdamsvej 17, 2100 Copenhagen, Denmark}
\author{Jose Mar\'ia Ezquiaga}  
\affiliation{Center of Gravity, Niels Bohr Institute, Blegdamsvej 17, 2100 Copenhagen, Denmark}
\author{Maya Silverman}
\affiliation{DARK, Niels Bohr Institute, University of Copenhagen, Jagtvej 155A, 2200 Copenhagen,  Denmark}
\author{Sarah Pearson}  
\affiliation{Technical University of Denmark, Elektrovej 327, 2800 Kgs. Lyngby, Denmark}
\affiliation{DARK, Niels Bohr Institute, University of Copenhagen, Jagtvej 155A, 2200 Copenhagen,  Denmark}

\begin{abstract}
The nature of dark matter remains elusive. 
A key open question is how much it clumps at small scales, 
with recent observations finding 
high concentrations consistent with core-collapsed subhalos. 
In this \this, we show that strongly lensed gravitational waves can discover or constrain such substructures, which would produce additional copies of the original signal and wave-optics features. 
We find that within 3 years of next-generation ground-based detectors, one could distinguish self-interacting scenarios from cold and warm dark matter at $\sim$ 3--5$\sigma$. 
Even with conservative rates, a single additional copy detected during 10 years of current observatories would disfavor cold dark matter at $>2\sigma$. 
Finally, we report that those collapsed low mass subhalos produce a median of $\sim5$ times greater wave-optics effects in future space-based detector's signals. This further motivates the use of gravitational waves as a premier probe of the nature of dark matter at small scales.

\end{abstract}


\maketitle

\lettersection{Introduction}\label{sec:intro}%
When looking into the distant past, every
cosmic messenger is sensitive to the distribution of matter in the Universe through gravitational lensing. 
In fact, this is how we collect some of the strongest evidence for the existence of \gls{DM}~\cite{Clowe:2006eq}. 
However, the resolution at which we
observe such lensed messengers determines the scale at which \gls{DM} substructures can be resolved. 
Importantly, it is only at small scales where currently viable \gls{DM} models differ from each other~\cite{Bullock:2017xww}. 

Electromagnetic observations of strongly lensed galaxies have proven a powerful \gls{DM} probe~\cite{Vegetti:2023mgp} and, in recent times, 
even hint at deviations from the standard \gls{CDM} scenario~\cite{Natarajan:2026ztd,Gilman:2026wul}. 
If confirmed, such observations point to the existence of very compact subhalos, which could be consistent with predictions from \gls{SIDM}~\cite{Adhikari:2022sbh}. 
On the opposite end, strong lensing also constrains 
\gls{WDM}, which suppresses the halo mass function~\cite{Gilman:2019nap,Nadler:2021dft,Keeley:2024brx}. 
Nonetheless, such surveys are intrinsically limited by their angular resolution to subhalo masses $\gtrsim10^6M_\odot$. 
Alternative methods, such as stellar streams or Milky Way satellite counts, offer complementary bounds in the same mass range~\cite{Banik:2019smi,DES:2020fxi}.

In the landscape of probing \dm substructures, 
\glspl{GW} appear as a promising new messenger. %
\Glspl{GW} are coherently emitted and detected with waveforms %
predicted from general relativity. 
Across cosmological propagation, these signals are only affected by gravitational lensing. %
The exquisite time resolution of current \gls{GW} detectors~\cite{LIGOScientific:2014pky,VIRGO:2014yos,KAGRA:2020tym}, down to the millisecond, implies direct sensitivity to lens masses as low as $\sim100M_\odot$. 
On the opposite end, the year-long observing runs guarantees the observation of lensing signatures up to the largest galaxy clusters $\sim10^{15}M_\odot$~\cite{Vujeva:2025kko, Shan:2026beq}. 
For masses up to $\sim10^5M_\odot$, \gls{GW} lensing in ground-based detectors is characterized by waveform distortions, as the repeated copies produced by strong lensing interfere and diffract with each other~\cite{Takahashi:2003ix}.
For higher lens masses, multiple copies arrive at the detector at different times, with distinct characteristics that can trace substructures~\cite{Vujeva:2025kko,Vujeva:2025nwg, Liu:2026onn, Swamy:2026ujz}. 

Approximately 1 in every 1000 \glspl{GW} from stellar-mass binaries is expected to be strongly lensed by galaxy-scale lenses and above~\cite{Ng:2017yiu,Oguri:2018muv,Xu:2021bfn,Wierda:2021upe,Smith:2022vbp,Li:2026dai}. 
This motivates current searches~\cite{LIGOScientific:2021izm,2023lensingO3full,GWTC4_lensing} and the exploration of \dm tests with future observations~\cite{Vujeva:2025nwg,Liu:2026onn}. 
Because the limited detection reach of current \gls{GW} observatories, strongly lensed signals are expected around regions of high-magnification~\cite{Lo:2024wqm}, 
where substructure effects are boosted~\cite{Vujeva:2025nwg}, and diffraction becomes relevant~\cite{Ezquiaga:2025gkd}. 
Importantly, any other astrophysical substructures, such as globular clusters or intermediate-mass black holes, are unlikely to produce observable effects in \glspl{GW} due to their low abundances near regions of high-magnification~\cite{Vujeva:2026xfh}. 
Future space-based detectors such as LISA are also promising probes, with larger subhalos leading to wave-optics features~\cite{Tambalo:2022wlm,Caliskan:2023zqm,Ando:2026poq,Ando:2026eam,Thakore:2026waw}. 

Given the promise of \gws as precision probes of masses currently out of reach to electromagnetic observatories, in this \this, we explore the constraining power that lensed \glspl{BBH} and \glspl{BNS} can have on the particle nature of \dm.
More specifically, we consider fiducial examples of the \cdm, \wdm, and \sidm paradigms, and quantify the distinctions we can make between instances of these models using current and next-generation detectors.

\begin{figure*}[!t]
    \centering
    \includegraphics[width=\linewidth]{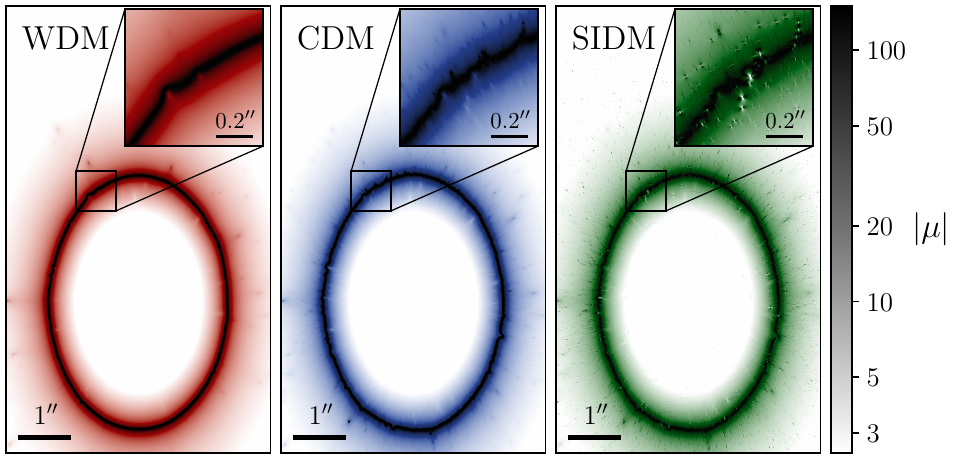}
    \caption{Image plane magnification maps for a $10^{13}M_\odot$ main halo at $z_{\rm L}=0.5$, and a source at $z_{\rm S}=2$ in \acrfull{WDM}, \acrfull{CDM} and \acrfull{SIDM}, from left to right.
    Subhalos with $\leq 10^7M_\odot$ in \gls{CDM}/\gls{WDM} produce small perturbations to the critical curves, whereas their core-collapsed equivalents in \gls{SIDM} can produce meaningful perturbations, as seen by the abundance of small-scale structure in the critical curves in the \gls{SIDM} panel.
    The \sidm example shown here is for the model with most collapsed subhalos considered in this work, SIDM100, with a viscosity cross section of $\sigma_\mathrm{v}/m = 100$ cm$^{2} \text{g}^{-1}$, to make the qualitative differences in the model clearly visible by eye.
    }
    \label{fig:image_plane_magnification_maps}
\end{figure*}

\lettersection{Modeling Dark Matter Subhalos}\label{sec:subhalo_model}%
The particle nature of \dm extensively changes the properties of both the density profile of each \dm subhalo, as well as their mass dependent number density~\cite{nadler_dark_2026}. 
In the standard \cdm paradigm, the density profiles of \dm subhalos are expected to follow the \tnfw profile~\cite{Navarro:1995iw, Navarro:1997mt}, whose density distribution is controlled by a characteristic mass $M$, and concentration $c$, truncation radius $r_\mathrm{t}$.
Thus, in our \cdm models, subhalos are simply described by \tnfw profiles~\cite{Navarro:1997mt}, with concentrations following the $c-M$ relation of Diemer and Joyce~\cite{Diemer:2018vmz}. 
For \wdm, the free-streaming of \dm particles suppresses both the abundance and concentrations of subhalos, particularly at low masses~\cite{Viel:2013fqw}, while their profiles remain \tnfw. This results in an under-abundance of \wdm subhalos at low masses, as well as less concentrated subhalos given the delay in subhalo formation due to the free-streaming at early times. In this work, we follow the prescription of Ref.~\cite{Benson:2026kdp}, and adopt a half-mode mass of $10^{7.5} M_\odot$.
\dm self-interactions instead reshape the subhalos themselves, first forming isothermal cores~\cite{Spergel:1999mh} and, given sufficient time, undergoing gravothermal core-collapse~\citep{Balberg:2002ue}.
These subhalos are modeled with a \tnfwc in order to capture the effects of the core formation.
Throughout this work, \dm %
self-interactions follow a velocity-dependent viscosity cross section $\sigma_\mathrm{v}(v)$ of the Rutherford-like form of Ref.~\cite{Yang2022Gravothermal}. To maximize the cross-section relevant for low-mass halos, which \glspl{GW} are sensitive to, while remaining consistent with existing constraints~\cite{Correa:2020qam, andrade_stringent_2021}, we set  the turnover velocity $w = 1\,\mathrm{km\,s^{-1}}$, while varying the characteristic cross-section, $\sigma_0/m$. Each model is labeled by its value at $v = 15\,\mathrm{km\,s^{-1}}$, namely $\sigma_\mathrm{v}/m = 10$, $50$, and $100\,\mathrm{cm^2\,g^{-1}}$. Henceforth, we refer to these models as SIDM10, SIDM50, and SIDM100, respectively.
Within this parametric model, increasing $\sigma_\mathrm{0}/m$ increases the collapse fraction of low-mass halos and increasing $w$ increases the subhalo masses that undergo collapse.
The core-collapse timescale $t_\mathrm{c}$ of each subhalo then follows from its \tnfwc parameters and velocity-averaged cross section~\cite{Yang:2023jwn}.
The lensing observables are sensitive to how far this collapse is allowed to proceed, since subhalo cores become increasingly compact as it continues.
In the parametric model of Ref.~\cite{Yang:2023jwn}, each subhalo's evolution is tracked through the dimensionless time $\tau = t/t_\mathrm{c}$.
While Ref.~\cite{Yang:2023jwn} stops the evolution of each halo at $\tau = 1$ to avoid extrapolating beyond the regime in which the model was calibrated, we find that this choice produces lensing observables indistinguishable to those of \cdm for our simulated observables. Instead, we evolve subhalos to $\tau \leq 1.6$, using analytic extrapolations of the scale and core radii, which produces substantially denser central regions.
Strong lensing analyses likewise model collapsed halos as deeply collapsed objects, with inner logarithmic slopes between $-2$ and $-3$~\cite{Gilman:2022ida}, or with profiles based on simulations that follow the collapse to $t/t_\mathrm{c} \sim 3$~\cite{Gilman:2026wul, Gurian:2025zpc}.
We also consider that tidal stripping accelerates core-collapse in subhalos relative to isolated halos~\cite{Nishikawa:2019lsc}. To account for this, we follow Ref.~\cite{Gilman:2022ida}, measuring halo ages from $z = 10$, and rescaling the subhalo collapse timescale as $t_\mathrm{c,sub} = \gamma_\mathrm{sub}\, t_\mathrm{c}$, with $\gamma_\mathrm{sub} = 0.4$.

Our choice is not intended to span the entire range of viable \sidm scenarios, nor to constrain them. Instead, it provides a representative example of a population of core-collapsed subhalos, which we use to illustrate the lensing signatures such a population can produce.
How to model subhalos at late stages of collapse remains a key open systematic in \sidm lensing studies~\cite{Gilman:2022ida, Yang:2024uqb}.
Even within this simple framework, the collapse history of the population depends on choices such as the turnover velocity and normalization of the cross section, and on $\gamma_\mathrm{sub}$, which sets how strongly tidal stripping accelerates collapse. Varying these quantities together leads to considerable degeneracies in the resulting subhalo population.
Beyond this framework lies a much broader landscape of \sidm models with additional particle physics, such as inelastic~\cite{vogelsberger_evaporating_2019} or dissipative self-interactions~\cite{roy_aggressively-dissipative_2024}, which we do not consider here.
We leave a systematic exploration of this parameter space, and of these alternative models, to future work.

The individual instances of the subhalo populations are generated using $\texttt{pyHalo}$~\cite{Gilman:2019nap, pyHalo:code}. 
In all cases, we consider a minimum mass of $10^5 M_\odot$, below which we find negligible contribution to the final results of this work. 
Finally, the subhalos are placed in a galaxy-scale lens, %
whose density distribution is given by an elliptical power law potential, which is characterized by a velocity dispersion, an axis ratio (taken to be 0.7 as a representative example), and a power law index, which was chosen to be 2, corresponding to an isothermal density distribution. 
The resulting image plane magnification maps for the three \dm models are presented in Fig.~\ref{fig:image_plane_magnification_maps}.

\begin{figure}[t!]
    \centering
    \includegraphics[width=\linewidth]{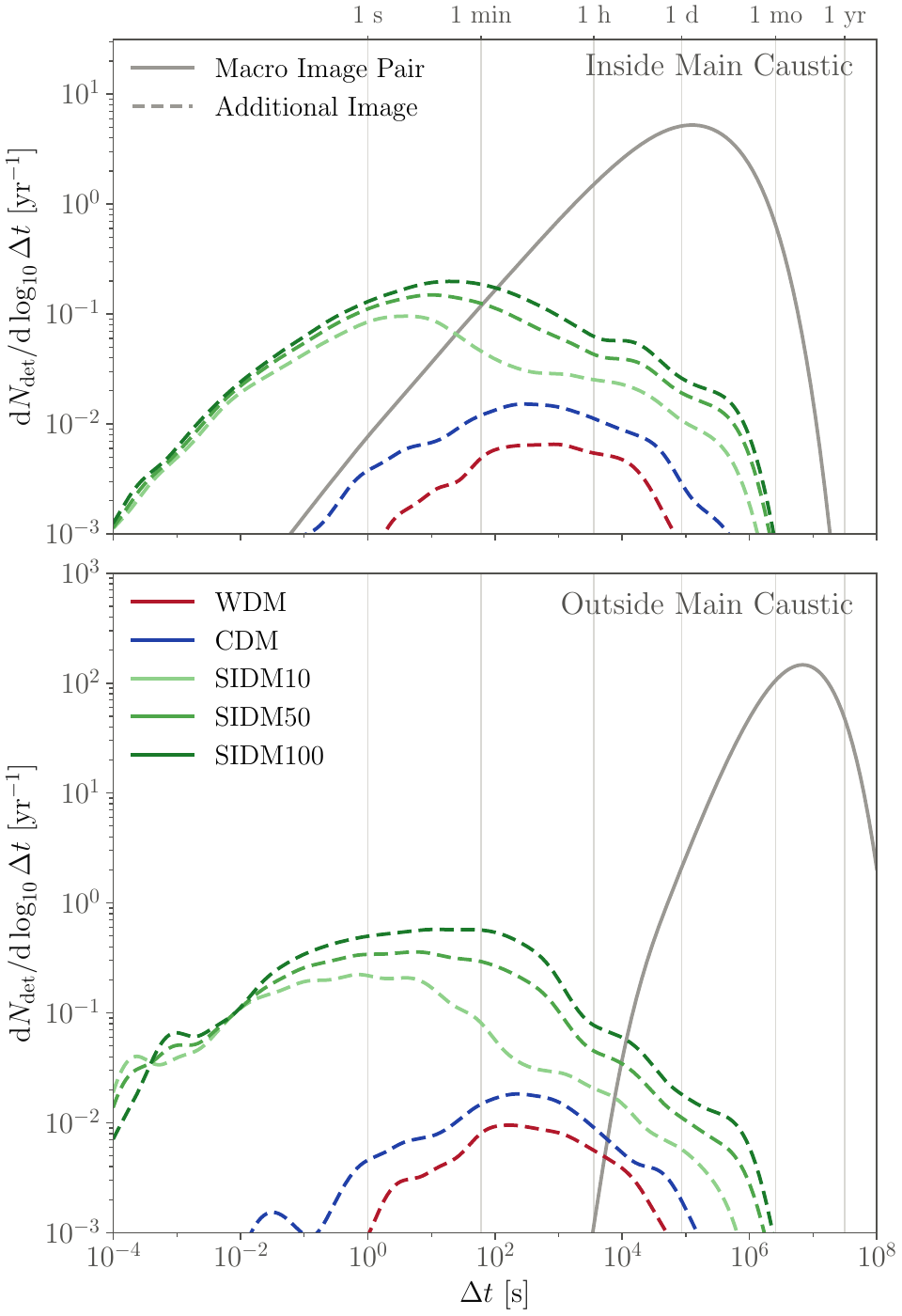}
    \caption{Time delay distribution of the shortest detectable image pairs with ground-based detectors (Einstein Telescope) inside (top figure) and outside (bottom figure) of the perturbed main halo caustics for \cdm, \wdm and \sidm, for the optimistic merger rate. 
    The production of additional structure in the caustics amplifies the high-magnification, and thus short time delay tail, for sources inside the main caustics in the \sidm models (green lines). 
    The core-collapsed halos production of additional caustics outside those of the main halos produces a much more pronounced population of short time delay image pair. These arise from the region outside of the main caustics, which in a smooth lens model, can generally only produce longer time delay pairs. 
    }
    \label{fig:timedelays}
\end{figure}

\lettersection{Gravitational Wave Lensing with Dark Matter Subhalos}\label{sec:sidm_lensing}%
Substructures in the potential of a galaxy introduce distinct observables in both the \go and \wo regimes. 
In \go, we have two main methods of inferring the existence of \dm substructure: 
through the appearance of additional highly magnified images, and through the relative magnification (i.e. the absolute value of the brightest image's magnification divided by that of the second brightest image) of highly magnified events exceeding a value of 2~\cite{Vujeva:2025nwg}. 
Due to the low redshift horizon of current detectors~\cite{LIGOScientific:2026sit}, 
and the steep increase with redshift of the optical depth, current (and near-future) ground-based detectors require relatively large magnifications, 
$\mu\gtrsim5$, for detection~\cite{Oguri:2018muv,Robertson:2020mfh}.
Since highly magnified images will inherently have shorter time delays~\cite{Schneider:1992, Lo:2024wqm, Baka:2025yqx, Rao:2025poe}, current detectors will tend to detect 
pairs and triplets of images with shorter time delays~\cite{Haris:2018vmn,More:2021kpb}. 
The increased sensitivity of future detectors will loosen this bias, capturing %
images produced by sources farther away from the caustics, as well as in the lower image multiplicity region outside of the main halo caustics that produces inherently lower magnifications, and longer time delays. 
In practice, a subhalo only alters images close to an existing image of the host. Therefore, we explicitly solve the system as a smooth host with subhalos nearby in small windows around individual host images: single images at fixed $|\mu|$ (minima, saddles and near-cusp images), and source boxes straddling a fold of the main caustic ($|\mu|\geq10$), whose statistics each host image then inherits.
In each window, we solve the lens equation for the host plus every subhalo within three Einstein radii ($\mathcal{O}(10^5)$), summing those near the images ($\mathcal{O}(10^3)$) exactly and the rest through a Chebyshev interpolant of their combined deflection. Image positions, magnifications and time delays agree with the full \texttt{lenstronomy}~\cite{Birrer:2018xgm} model to $10^{-8}$\,arcsec, $10^{-5}$ and $10^{-5}$, respectively, at more than $10^{3}$ times lower cost.
An event has an additional image when a host image splits into three or more images and the second brightest of that group (the third, for a fold pair) is detected; $R=|\mu_1/\mu_2|$ is computed from the two brightest images of the group.
For each of the host masses and dark matter models, we use 300 subhalo populations (20 at each of 15 lens redshifts up to $z_l=3$), each lensing 200 sources per source redshift out to $z_s=20$, giving more than $10^{5}$ lens simulations in total.

The time delay distribution of the shortest detectable image pair provides valuable insights, as seen in Fig.~\ref{fig:timedelays}. 
For source positions \emph{inside} the main caustic, \sidm subhalos enhance the short time delay tail below arrival time differences of 10 seconds
(corresponding to an effective lens mass of $\sim 10^6M_\odot$). 
This enhancement is much more significant than in \cdm, and almost non-existent in \wdm. 
There, the short time delay tail is well below what is predicted by a smooth lensing potential (solid line).
We have verified that the tail is preserved when the minimum subhalo mass is increased to $10^6M_\odot$ the shape of the tail is preserved, while when increasing it to $10^7M_\odot$, it disappears. 
At the other end, lowering the minimum mass to $10^4M_\odot$ does not change the break point or enhancement, indicating that it is the $10^6M_\odot$ halos which dominate the effect. 
For sources \emph{outside} the main caustic, 
short time delays are solely produced by the subhalos, with \sidm again boosting the rate compared to \cdm and \wdm. 
In fact, with future ground-based detectors, most of the new detectable images lie outside the main caustic, as shown in Fig.~\ref{fig:timedelays}. 
This is due to their increased sensitivity to images with lower magnifications, which dominate the cross section of strong lensing.
In contrast, new signals in current detectors are dominated by subhalos within the main caustic, where \sidm still exceeds \cdm, but by a lesser amount.
For comparison, with current detectors, typical \glspl{BBH} last for up to $\sim$seconds in band, while \glspl{BNS} were observed for more than a minute. 
Therefore, the repeated new copies produced by subhalos could lead to overlapping \glspl{GW}~\footnote{Recent studies have claimed to have not found additional images formed by CDM subhalos~\cite{Swamy:2026ujz}. However, this appears to be due to insufficient sampling of the source plane. Our rate of additional image formation in CDM is substantially lower than their 200 simulated events.}.

In the \wo regime, we solve the diffraction integral numerically using \glow~\cite{Villarrubia-Rojo:2024xcj, GLoW:code}.
We quantify the effect of subhalos by computing the frequency dependent variations of the diffraction integral compared to a case without them 
(i.e. the maximum of $|F(f)/F_{\rm smooth}(f)-1|$ across the band, normalized at its highest frequency so that the achromatic change in magnification is removed). 
For ground-based detector sources, defined as $f_{\rm min}>10$Hz, we find that the Fresnel region is so small, $r_{\rm F}\simeq[cD_{\rm l}D_{\rm ls}/(\pi(1+z_{\rm l})D_{\rm s}f|\lambda|)]^{1/2}\approx0.4\,{\rm pc}\,(f/10\,{\rm Hz})^{-1/2}|\lambda|^{-1/2}$ (i.e. $\sim1$\,pc for a typical image with $\mu\sim5$), that the number density of subhalos is not high enough to give an effect: on average only $\sim10^{-2}$ subhalos above $10^3M_\odot$ fall within it, and the typical image varies by $\lesssim10^{-4}$ in $|F|$ (mismatches $\lesssim10^{-7}$). 
Only rare close encounters ($\lesssim1$ percent of images) with collapsed \sidm subhalos produce $\sim10\%$ features. 
For space-based detector sources, defined as $10^{-3}\leq f\leq10^{-1}$Hz, we find variations of $\sim0.6\%$ in \cdm (0.2--2 percent across the 5th--95th percentiles), in agreement with~\cite{Ando:2026poq}. 
When moving to \sidm, these effects are boosted by a median of $\sim$5--6 times, with \sidm exceeding \cdm in 96\% of sight-lines. 
On the contrary, \wdm leads to negligible \wo features. 
We do not consider the effects of lens stochastic diffraction that may be extracted in a population level analysis~\cite{Choi:2026lsa}.

\begin{figure}[!t]
    \centering
    \includegraphics[width=\linewidth]{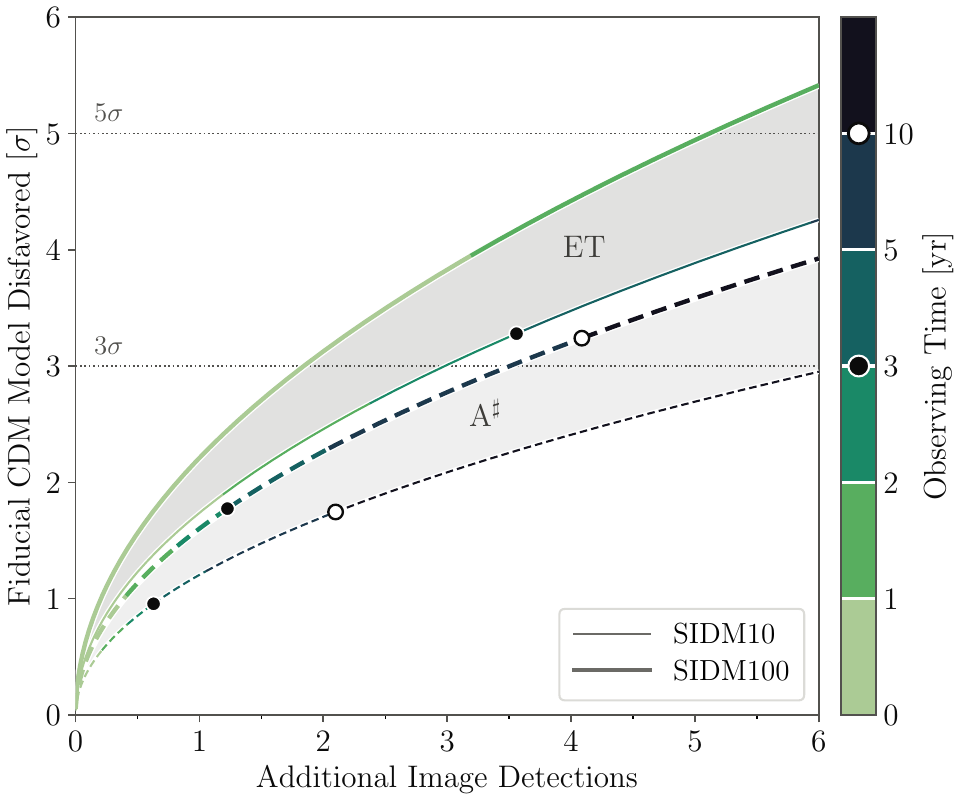}
    \caption{The significance at which  
    our fiducial \cdm model would be disfavored given the additional new detectable images produced by \sidm subhalos, 
    shown for our lowest and highest viscosity cross sections SIDM10, and SIDM100, with thin and thick lines respectively. 
    We forecast constraints for LIGO's planned upgrade (A$^\sharp$, dashed) and next-generation detector Einstein Telescope (ET, solid) for an optimistic \gls{BBH} merger rate consistent with GWTC-5. 
    The color bar tracks the observing time, with the filled and unfilled black circles highlighting 3 and 10 years.
     }
    \label{fig:constraints}
\end{figure}

\lettersection{Constraints on dark matter}\label{sec:constraints}%
The presence of compact \dm subhalos leaves distinct signatures in strongly lensed \glspl{GW}. 
Naturally, the (non)observation of such effect allows us to constrain the underlying \dm model. 
For \sidm, we control the fraction of core-collapsed halos through the viscosity cross section over mass, $\sigma_{\rm V}/m$, which varying from 10 to 100 $\mathrm{cm}^2/{\rm g}$ changes the fraction of collapsed low-mass subhalos.
A larger $\sigma_{\rm V}/m$, results in more higher mass subhalos collapsing. 
As a consequence, we find that all the observable properties of subhalo lensing (high image multiplicity, large relative magnifications, short time delays, and wave-optics features) increase with $\sigma_{\rm V}/m$.
In order to forecast constraints on \sidm with ground-based detectors, we consider a population of stellar-mass compact objects (\glspl{BBH} and \glspl{BNS}) as sources, whose properties are consistent with the latest observations from \gls{LVK}~\cite{LIGOScientific:2026ctl} and whose merger rate is extrapolated at high redshift to follow the star formation rate~\footnote{More specifically, we choose values of the merger-rate density $\mathcal{R}(z)\propto(1+z)^{\alpha_z}/[1+((1+z)/(1+z_p))^{\alpha_z+\beta_z}]$ with a local rate at the upper end of the GWTC-5 90\% interval, $\mathcal{R}(z=0.2)=36.85\,{\rm Gpc^{-3}\,yr^{-1}}$, and a redshift evolution that peaks later and falls off more slowly ($\alpha_z=3.2$, $z_p=3$, $\beta_z=2.25$) than the GWTC-5 fiducial model~\cite{LIGOScientific:2026ctl} ($\mathcal{R}(z=0.2)=30.4\,{\rm Gpc^{-3}\,yr^{-1}}$, $\alpha_z=2.54$, $z_p=1.9$, $\beta_z=3.06$). }. 
We also assume standard distributions for the main halos, which are drawn from a Schechter function fitted to the observations of the Sloan Digital Sky Survey~\citep{Choi:2006qg}, for galaxy masses of $10^{10} - 10^{13}M_\odot$ at redshifts of $z_\mathrm{L}\in[0.1,3.0]$.
Sources are placed behind each simulated lens at $z_\mathrm{S}\in[z_\mathrm{L} + 0.1,20.0]$. 
We consider different ground-based detector configurations ranging from \gls{LVK} at present sensitivity (O4), design sensitivity (O5)~\cite{sensitivity_curves_ligo}, LIGO's planned upgrade (A$^\sharp$)~\cite{KAGRA:2013rdx,LIGO:2026gzb}, and next-generation detector Einstein Telescope (ET)~\cite{ET:2019dnz}. 
Similar results hold for Cosmic Explorer~\cite{Evans:2021gyd}.

We find that current detectors are similarly sensitive to the effects of subhalos in high-magnification regions as future detectors.
Therefore, the first strongly lensed \gls{GW} observations with \gls{LVK} are proportionally more likely to have additional images from subhalos, ranging from 4.1\% to 6.5\%, from SIDM10 to SIDM100 respectively.
For reference, our fiducial \cdm model predicts 1.5\%.
With improved sensitivity in the detectors, more strongly lensed \glspl{GW} are detected in general.
The gain in subhalo signatures is not proportional,
as these events typically come from regions with lower magnifications: from present O4 sensitivity to ET, the strongly lensed rate grows by a factor of $\sim$600, the additional-image rate by only $\sim$60--100, and the fraction falls to 0.4--1\%.
Importantly, we find that in $\sim$45\% of cases the new detectable copies in O5 have time delays shorter than 1 second, potentially leading to overlapping signals.

Figure~\ref{fig:constraints} summarizes how well A$^\sharp$ and ET can distinguish between \dm models.
As only a small fraction of lensed events produce additional images, this constraining power depends strongly on the \gls{GW} merger rate, particularly at high redshifts where lensing is most efficient.
We stress that these constraints apply only to our fiducial \cdm and \wdm models described above, rather than to these paradigms in general.
With optimistic merger rates, ET could discern our fiducial \cdm model from \sidm, even for the mildest cross section (SIDM10), at 3$\sigma$ within 2.5 years, and from SIDM100 at 5$\sigma$ within only 1.6 years.
Even A$^\sharp$ could see first hints of \sidm subhalos after ten years of observations, differentiating our fiducial \cdm model at 1.7, 3.2$\sigma$ from SIDM10 and SIDM100 respectively.
With conservative merger rates, constraints become much harder to obtain: after ten years of ET observations, our fiducial \cdm model could be distinguished at only 3.4$\sigma$, even for the largest cross section, SIDM100.
Our fiducial \wdm model, in contrast, predicts additional-image rates approximately half that of \cdm, and thus remains indistinguishable from it on these timescales using the additional detectable image probe considered in this work.

Based on current merger rates~\cite{LIGOScientific:2025pvj}, we find that strongly lensed \glspl{BNS} are unlikely until next-generation detectors. 
To be observable by current detectors, very high magnifications are required, $\mu\gtrsim100$~\cite{Smith:2022vbp}.
We find that such detections are more likely to be affected by subhalos, with a probability that rises to 25-30\%.

\lettersection{Conclusions}\label{sec:conclusions}
The nature of \dm is still elusive
and the key to discern between different paradigms could lie in the small scale cosmological structures. 
We demonstrate that strongly lensed \glspl{GW} are sensitive to \dm subhalos, with effects growing rapidly with their compactness. 
Even if \dm only self-interacts with the viscosity cross section of SIDM10, future ground-based detectors could disfavor our fiducial \cdm model at 3$\sigma$ within 2.5 years for optimistic merger rates, while SIDM100 would allow it to be excluded at 5$\sigma$ within only 1.6 years.
Even A$^\sharp$ could see first hints of \sidm subhalos after ten years of observations.
With conservative merger rates, ten years of observations would disfavor our fiducial \cdm and \wdm models at only $\sim$3$\sigma$ and $\sim$4$\sigma$, respectively, even for the largest cross section.

The existence of core-collapsed subhalos in \sidm-like scenarios predicts enhanced \wo signatures in the \glspl{GW} detected by LISA~\cite{LISACosmologyWorkingGroup:2022jok,Tambalo:2022wlm,Caliskan:2023zqm}. 
There, the richness of the frequency dependent waveform distortions could bring new insights into the density profile of such subhalos. 
Additional work is needed, however, into solving and characterizing these diffraction patterns, 
as well as implementing them in search pipelines~\cite{Chan:2024qmb,Chan:2026alt,Wadekar:2026xse}. 
In future work, we will explore the broader range of \sidm models, from alternative cross sections and collapse prescriptions, to additional dark-sector physics~\cite{Tulin:2017ara, Adhikari:2022sbh}, and how these choices affect the lensing observables and the constraints they place on \sidm.
Similarly, more work is required to discern the impact of \dm subhalos for multi-messenger lensing searches~\cite{Smith:2025axx}, as we have found large probabilities of additional images in highly magnified \glspl{BNS}.
In those high-magnification cases, stellar fields could also contribute with additional distortions~\cite{Mishra:2021xzz,Shan:2022xfx}.

While detailed modeling of the cosmic formation history of \dm subhalos still needs to be accounted for, as well as the uncertainties in the merger rate of compact binaries that propagate into the lensing rates, our work shows that strongly lensed \glspl{GW} are unique probes of \dm substructures and can constrain the amount of core-collapsed subhalos, favoring or ruling out competing \dm scenarios.

\emph{Acknowledgments}.---We are grateful to Mark Cheung for useful conversations, and to Sreekanth Harikumar for reviewing the paper internally in LVK. 
Claude Opus 5.5 was used to assist in coding and verifying grammar in the manuscript. LV carefully reviewed all of Claude's outputs and takes full responsibility for all code used in this work. All scientific conclusions are the authors’ own.
This work was supported by Research Grants No.~VIL37766 and No.~VIL53101 from Villum Fonden and the DNRF Chair Program Grant No.~DNRF162 by the Danish National Research Foundation.  
J.~M.~E is also supported by the Marie Sklodowska-Curie Grant Agreement No.~847523 INTERACTIONS.
%
The Center of Gravity is a Center of Excellence funded by the Danish National Research Foundation under grant no.~DNRF184.
This project has received funding from the European Union's Horizon 2020 research and innovation programme under the Marie Sklodowska-Curie grant agreement No 101299389 and No 101131233.
This work is supported by Simons Foundation International \cite{sfi} and the Simons Foundation \cite{sf} through Simons Foundation grant SFI-MPS-BH-00012593-11. 
MS and SP were supported by a research grant (VIL53081) from VILLUM FONDEN.
Co-funded by the European Union (ERC, BeyondSTREAMS, 101115754). Views and opinions expressed are however those of the author(s) only and do not necessarily reflect those of the European Union or the European Research Council. Neither the European Union nor the granting authority can be held responsible for them.
The Tycho supercomputer hosted at the SCIENCE HPC center at the University of Copenhagen was used for supporting this work. 
The authors also acknowledge the use of the Sophia workstation for this work.

\bibliography{gw_lensing_fixed}
\appendix

\end{document}

%% file: acronyms.tex
\usepackage[acronym]{glossaries}
\newacronym{GW}{GW}{gravitational wave}
\newacronym{EM}{EM}{electromagnetic} 
\newacronym{GL}{GL}{gravitational lensing}
\newacronym{GO}{GO}{geometric optics} 
\newacronym{WO}{WO}{wave optics}
\newacronym{CBC}{CBC}{compact binary coalescence}
\newacronym{BBH}{BBH}{binary black hole}
\newacronym{BNS}{BNS}{binary neutron star}
\newacronym{NSBH}{NSBH}{neutron-star black-hole binary}
\newacronym{SIE}{SIE}{singular isothermal ellipsoid}
\newacronym{SNR}{SNR}{signal-to-noise ratio}
\newacronym{PE}{PE}{parameter estimation}
\newacronym{LVK}{LVK}{LIGO--Virgo--KAGRA}
\newacronym{CDM}{CDM}{cold dark matter}
\newacronym{WDM}{WDM}{warm dark matter}
\newacronym{SIDM}{SIDM}{self-interacting dark matter}
\newacronym{DM}{DM}{dark matter}
\newacronym{NFW}{NFW}{Navarro-Frenk-White}
\newacronym{TNFW}{TNFW}{truncated Navarro-Frenk-White}
\newacronym{TNFWC}{TNFWC}{cored truncated Navarro-Frenk-White}